\documentclass[british,english,aps, reprint, prl,twocolumn]{revtex4}
\usepackage[T1]{fontenc}
\usepackage[latin9]{inputenc}
\usepackage{color}
\usepackage{textcomp}
\usepackage{amssymb}
\usepackage{graphicx}

\makeatletter

\providecommand{\tabularnewline}{\\}

\@ifundefined{textcolor}{}
{%
 \definecolor{BLACK}{gray}{0}
 \definecolor{WHITE}{gray}{1}
 \definecolor{RED}{rgb}{1,0,0}
 \definecolor{GREEN}{rgb}{0,1,0}
 \definecolor{BLUE}{rgb}{0,0,1}
 \definecolor{CYAN}{cmyk}{1,0,0,0}
 \definecolor{MAGENTA}{cmyk}{0,1,0,0}
 \definecolor{YELLOW}{cmyk}{0,0,1,0}
}

\makeatother

\usepackage{babel}
\begin{document}

\title{Flux Qubits with Long Coherence Times for Hybrid Quantum Circuits}

\author{M. \surname{Stern}$^{1}$}

\email{michael.stern@cea.fr}

\author{G. \surname{Catelani}$^{2}$}

\author{Y. \surname{Kubo}$^{1}$}

\author{C. \surname{Grezes}$^{1}$}

\author{A. \surname{Bienfait}$^{1}$}

\author{D. \surname{Vion}$^{1}$}

\author{D. \surname{Esteve}$^{1}$}

\author{P. \surname{Bertet}$^{1}$}

\affiliation{$^{1}$ Quantronics Group, SPEC, IRAMIS, DSM, CEA Saclay, Gif-sur-Yvette,
France}

\affiliation{$^{2}$Forschungszentrum Jülich, Peter Grünberg Institut (PGI-2),
52425 Jülich, Germany}
\begin{abstract}
We present measurements of superconducting flux qubits embedded in
a three dimensional copper cavity. The qubits are fabricated on a
sapphire substrate and are measured by coupling them inductively to
an on-chip superconducting resonator located in the middle of the
cavity. At their flux-insensitive point, all measured qubits reach
an intrinsic energy relaxation time in the 6-20 \textmu{}s range and
a pure dephasing time comprised between 3 and 10 \textmu{}s. This
significant improvement over previous works opens the way to the coherent
coupling of a flux-qubit to individual spins.
\end{abstract}
\maketitle

Electronic spins in semiconductors such as NV centers in diamond or
phosphorus donors in silicon can reach coherence times up to seconds
\cite{Bar-Gill,Morello,Saeedi} and are therefore promising candidates
for quantum information processing. However, the main obstacle to
an operational spin-qubit quantum processor is the difficulty of coupling
distant spins. To solve this issue, it has been proposed to couple
each spin to a superconducting circuit which acts as a quantum bus
and mediates the spin-spin interaction \cite{Marcos,Twamley}. This
approach requires to reach the strong coupling regime where the coupling
strength $g$ between the spin and the circuit is larger than their
respective decoherence rates. Among all superconducting circuits,
the largest coupling constants could be obtained with flux qubits
(FQ) \cite{Mooij,Orlando,vanderWal,Chiorescu,Bertet,Yoshihara,Forn-Diaz,Bylander}
due to their large magnetic dipole. They can reach up to $g/2\pi\sim100$~kHz
for realistic parameters, which is much larger than the best reported
spin decoherence rates. This brings a strong motivation for developing
FQs with a coherence time $T_{2}>2/g\sim4$~$\mu\mathrm{\mathrm{s}}$.

FQ coherence times reported up to now are limited to $T_{2}\lesssim2\mu\mathrm{s}$,
with a sizeable irreproducibility \cite{Bertet,Bylander}. The reasons
for these relatively short coherence times are numerous but stem in
part from the poor control of the electromagnetic environnement of
the qubit in previously used dc-SQUID readout setups \cite{Chiorescu,Bertet,Yoshihara}\textcolor{black}{.
A better control of the environnement was recently demonstrated in
the case of another superconducting qubit, the transmon, by using
a three dimensional (3D) cavity that allows reading out the qubit
and protecting it from spontaneous emission \cite{Paik}. }A natural
question is therefore whether it is also possible to increase the
coherence times of FQs and their reproducibility by using such a setup. 

In this work, we present the first measurements of FQs in a 3D cavity.
The six qubits measured reach reproducible coherence times $T_{2}$
between $2$ and $8\,\mu\mathrm{s}$, which would be already sufficient
to reach the strong coupling regime with a single spin. In addition,
our results shed light on decoherence of FQs, giving evidence that
charge noise is the dominant decoherence mechanism at their flux-insensitive
point.

\begin{figure}
\includegraphics[width=1\columnwidth]{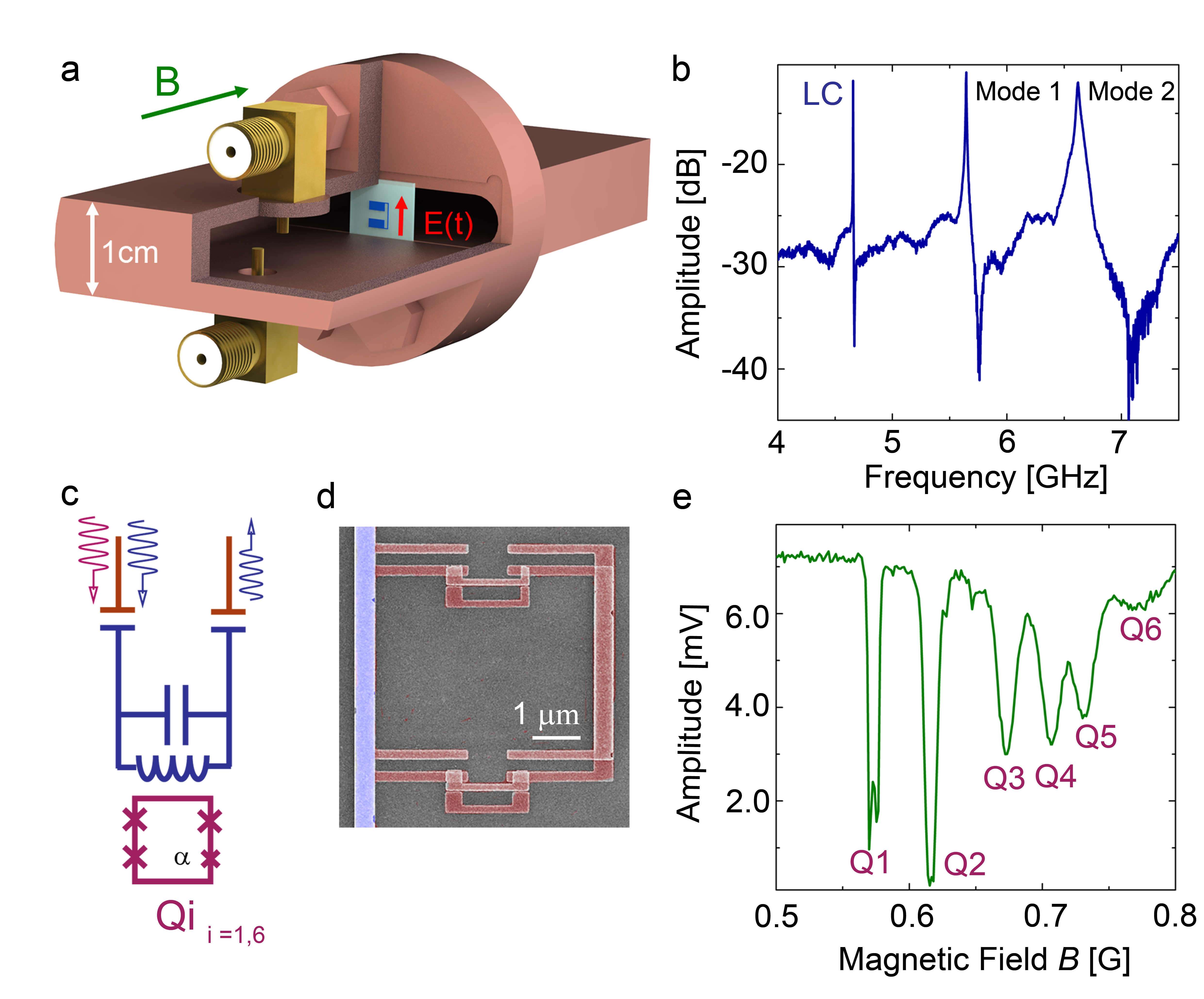}\caption{(a) Cut-away representation of the 3D cavity, with the LC circuit
(in blue) on its sapphire chip. The green arrow represents the applied
magnetic field $B$. The red arrow represents the ac electric field
$E(t)$ of the first mode of the cavity. (b) Transmission spectrum
of the cavity coupled to the LC resonator. The first peak at frequency
$\omega_{LC}/2\pi=4.643$ GHz corresponds to the resonance of the
LC resonator while the two other peaks correspond to the first modes
of the cavity. (c-d) Circuit diagram and colorized SEM micrograph
showing the FQ (in red) inductively coupled to the LC resonator (in
blue). (e) Amplitude of the transmitted signal at frequency $\omega_{LC}$
as a function of $B$, showing the signal from the six qubits.}
\end{figure}

\begin{figure*}
\includegraphics[width=1\textwidth]{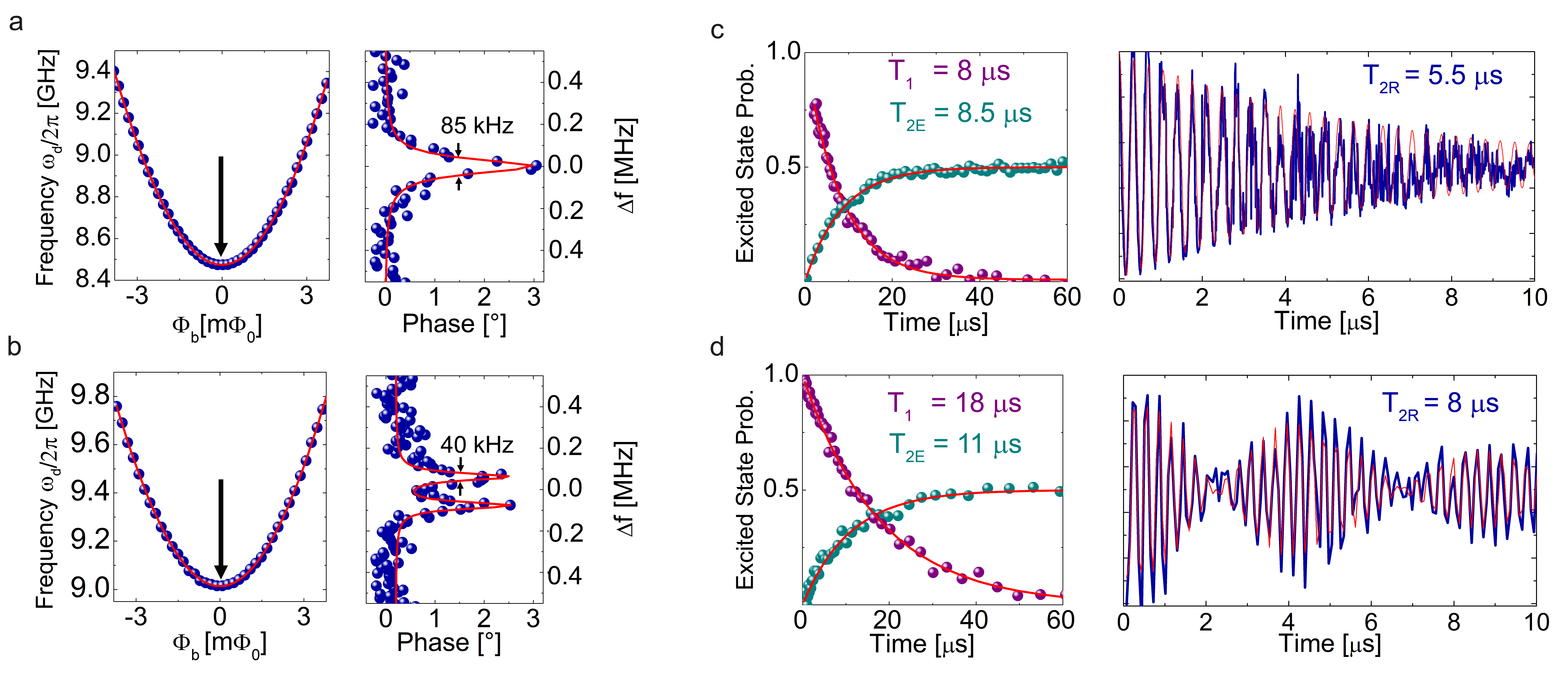}\caption{Characterisation of $Q3$ (top panels) and $Q4$ (bottom panels):
(a-b) (left panels) Measured qubit frequency $\omega_{01}(\Phi_{b})$
(blue dots) and fit (red curve) yielding the qubit parameters $\Delta$
and $I_{P}$. (Right panels) Spectroscopy data at $\Phi_{b}=0$ (blue
dots). $Q3$ spectrum is fitted with a single Lorentzian peak ; $Q4$
spectrum is fitted with a sum of two Lorentzian peaks separated by
150kHz. (c-d) (left panels) Qubit energy relaxation and spin-echo
measurements. The excited state probability is plotted as a function
of the delay between the $\pi$ pulse and readout pulse (blue dots)
or between the two $\pi/2$ pulses of the echo sequence (purple dots).
Red (orange) solid line is an exponential fit to the energy relaxation
(spin-echo) data. (Right panels) Measured Ramsey fringes (blue solid
line), with fit (red solid line) to exponentially damped single (top)
and double (bottom) sine curves. }
\end{figure*}

A scheme of the three dimensional cavity used in our experiment is
shown in Fig.~1a. The cavity is made of copper to enable the application
of an external magnetic field $B$ to the FQs. Its dimensions are
chosen for its first mode to be at 5.6 GHz. The sample inserted in
the cavity is a sapphire chip with an LC resonator inductively coupled
to six FQs, with a coupling constant $\sim50$ MHz. The LC resonator
acts as an intermediate coupler \cite{Pop} between the FQs and the
first cavity mode. It appears as a resonance peak at $\omega_{LC}/2\pi=4.643$
GHz in the transmission spectrum (Fig.~1b) with a quality factor
$Q_{LC}\sim1.5\times10^{4}$ determined by the length of the input
and output antennas inside the cavity (see Fig.~1a). 

Figure 1d presents a colorized SEM micrograph of one of the FQs. It
consists of a superconducting aluminum loop of area $A$ intersected
by four Josephson junctions. Three of the junctions are identical
with a Josephson energy $E_{J}/h=250$~GHz and a single electron
charging energy $E_{C}/h=3.6$~GHz; the fourth junction area is smaller
than others by a factor $\alpha$ (see Table I). When the flux threading
the loop $\Phi=B\, A$ is close to half a flux quantum $\Phi_{0}/2$,
the two states characterized by a persistent current $\pm I_{P}$
in the loop become degenerate, hybridise and give rise to an energy
splitting $\hbar\Delta$ called the flux-qubit gap. This circuit behaves
therefore as a two-level system and its transition frequency is $\omega_{01}=\sqrt{\Delta^{2}+\varepsilon^{2}}$
with $\varepsilon=2\, I_{P}\Phi_{b}/\hbar$ and $\Phi_{b}=\Phi-\Phi_{0}/2$
\cite{Mooij,Orlando}. 

The qubits are fabricated by double angle-evaporation of Al\textendash{}AlOx\textendash{}Al
on sapphire . We use a tri-layer PMMA-Ge-MAA process \cite{Romijn},
which provides a good precision and reproducibility of the junction
size and a rigid germanium mask, robust to the $\mathrm{O}_{2}$ ashing
and ion milling cleaning steps, which evacuates efficiently the charges
during e-beam lithography. The measurements are performed in a cryogen-free
dilution refrigerator at a temperature of 33 mK. The device is magnetically
shielded with 2 Cryoperm boxes surrounding a superconducting enclosure.
The cavity is also closed using Ecosorb corks and seals, in order
to protect the sample from electromagnetic radiation that could generate
quasiparticles \cite{BarendsAPL}. The input line is attenuated at
low temperature to minimize thermal noise and filtered with impedance-matched
radiation-absorbing filters. The readout output line includes several
filters, two isolators and a cryogenic HEMT amplifier. Qubit state
manipulations are performed by injecting in the input line microwave
pulses at $\omega_{d}\sim\omega_{01}$, followed by a readout pulse
at $\omega_{m}\sim\omega_{LC}$ whose amplitude and phase yield the
qubit excited state probability \cite{Blais}.

An advantage of the cavity readout is the possibility to measure several
qubits in a single run, by fabricating them with different loop areas
$A_{i}$ so that the field $B_{i}=\Phi_{0}/2A_{i}$ at which the flux
reaches $\Phi_{0}/2$ is different for each qubit. Figure 1e presents
the amplitude of the transmitted signal at frequency $\omega_{LC}$
as a function of $B$, showing a dip in the amplitude of the transmitted
signal when the frequency of any of the six FQs comes close to $\omega_{LC}$.

Each qubit is characterized by its spectroscopic parameters $\Delta$
and $I_{P}$, extracted from the dependence of its resonance frequency
on the applied flux. These values, given in Table 1, are in good agreement
with the predictions of the model described in \cite{Supplementary}
using both the measured values of $\alpha$ and of the tunnel resistance
of the junctions. The coherence properties of each qubit are measured
with the appropriate microwave pulse sequence \cite{Vion}: the energy-relaxation
time $T_{1}$, the Ramsey coherence time $T_{2R}$ from which one
gets the Ramsey pure dephasing time $(T_{\varphi R})^{-1}=(T_{2R})^{-1}-(2T_{1})^{-1}$,
and the echo decay time $T_{2E}$ yielding the echo pure dephasing
time $(T_{\varphi E})^{-1}=(T_{2E})^{-1}-(2T_{1})^{-1}$. 

We now present detailed measurements on the qubits $Q3$ and $Q4$
having the longest coherence times. The flux dependence of their frequency,
shown in Fig.2a (b), yields $\Delta/2\pi=8.47$~($9.01)$~GHz and
$I_{P}=169$ ($160$)~nA for Q3 (Q4). Since both qubits were designed
to have the same parameters, this shows good control of our e-beam
lithography and oxidation parameters. We now turn to the coherence
times at the so-called optimal point $\Phi_{b}=0$, where the qubit
frequency $\omega_{01}=\Delta$ is insensitive to first order to flux-noise
\cite{Bertet,Yoshihara}. Energy relaxation (see Fig.2c-d) is exponential
with $T_{1}=8\,\mu\mathrm{s}$ for $Q3$ and $18\,\mu\mathrm{\mathrm{s}}$
for $Q4$. Ramsey fringes also show an exponential decay for $Q3$
with $T_{\varphi R}=8.5\,\mu\mathrm{s}$, and an exponentially decaying
beating pattern for $Q4$ with $T_{\varphi R}=10\,\mu\mathrm{s}$.
These features are consistent with the qubit spectra measured after
an excitation pulse of $\sim20\:\mu\mathrm{s}$ with a power well
below saturation (see Fig.2a): $Q3$ line is Lorentzian with a full-width-half-maximum
(FWHM) of $85$ kHz, while $Q4$ line consists of a doublet of two
Lorentzians separated by $150$ kHz and having a width of $40$ kHz,
whose origin is discussed further below. 

The amplitude of the spin-echo signal decays exponentially (see Fig.2
c,d) with finite pure dephasing times $T_{\varphi E}=17\,\mu\mathrm{s}$
for $Q3$ and $T_{\varphi E}=16\,\mu\mathrm{s}$ for $Q4$. This moderate
improvement compared to the Ramsey pure dephasing times points out
to the presence of high-frequency noise in our circuit in contrast
to previous reports \cite{Bylander,Yoshihara}. We attribute this
effect to fluctuations in the photon number (photon noise) of one
or several cavity modes inducing fluctuations of the qubit frequency
due to the dispersive shift \cite{Bertet,Rigetti,Sears}. This noise
cannot be compensated by the echo protocol because its correlation
time ($\sim100$~ns - $1\mu\mathrm{s}$) given by the mode energy
damping is shorter than the echo sequence duration. By plunging the
antennas deeper in the cavity for reducing the cavity damping time,
the observation of a lower $T_{2E}$ confirms this explanation \cite{Rigetti,Sears}.
Interestingly, removing the photon-noise contribution from the Ramsey
pure dephasing time yields a ``low-frequency Ramsey dephasing time''
$(\tilde{T}_{\varphi R})^{-1}=(T_{\varphi R})^{-1}-(T_{\varphi E})^{-1}$
with $\tilde{T}_{\varphi R}=16\,\mu\mathrm{s}$ for $Q3$ and $\tilde{T}_{\varphi R}=30\,\mu\mathrm{s}$
for $Q4$. This one order of magnitude improvement compared to previous
flux-qubit experiments that reported $\tilde{T}_{\varphi R}$ in the
$0.2-2.5\mu\mathrm{s}$ range at the optimal point \cite{Yoshihara,Bertet,Bylander}
is discussed later.

\begin{table}[b]
\selectlanguage{british}%
\resizebox{\columnwidth}{!}{%

\selectlanguage{english}%
\begin{tabular}{|c|c|c|c|c|c|c|c|}
\hline 
 & $\Delta/2\pi$ (GHz) & $I_{P}$(nA) & $\alpha$ & $T_{1}$$(\mu\mathrm{s})$ & $T_{P}$$(\mu\mathrm{s})$ & $T_{\varphi R}$$(\mu\mathrm{s})$ & $T_{\varphi E}$$(\mu\mathrm{s})$\tabularnewline
\hline 
Q1 & 2.70 & 245 & 0.61 & 6-10 & $1.1\times10^{5}$ & 2 & 7\tabularnewline
Q2 & 4.91 & 207 & 0.55 & 2 & $3$ & - & -\tabularnewline
Q3 & 8.47 & 169 & 0.49 & 6.5-8 & $30$ & 8 & 17\tabularnewline
Q4 & 9.01 & 160 & 0.49 & 13-18 & 270 & 10 & 16\tabularnewline
Q5 & 9.71 & 171 & 0.43 & 5.5-12 & 90 & 5 & >100\tabularnewline
Q6 & 15.15 & 140 & 0.4 & 4 & 12 & - & -\tabularnewline
\hline 
\end{tabular}}

\caption{Parameters of the different measured FQs. Here $\Delta$ is the FQ
gap, $I_{P}$ is the persistent current flowing in the loop of the
qubit, $\alpha$ corresponds to the measured ratio between the small
and big junctions, $T_{1}$ is the relaxation time, $T_{P}$ the Purcell
limit time due to the coupling of the qubit with the cavity, $T_{\varphi R}$
and $T_{\varphi E}$ the coherence times obtained by Ramsey and Echo
sequences respectively.}
\end{table}

Away from the optimal point, decoherence is governed mainly by flux-noise
in agreement with previous works \cite{Yoshihara,Bylander}. The Ramsey
and spin-echo damping become Gaussian as $|\Phi_{b}|$ increases,
which is characteristic of $1/f$ noise \cite{Ithier}. Fitting the
Ramsey (or echo) envelope as $f_{R,E}(t)=e^{-t/(2T_{1})}e^{-(\Gamma_{\varphi R,E}t)^{2}}$,
we observe a linear dependence of $\Gamma_{\varphi R,E}=(T_{\varphi R,E})^{-1}$
on $|\Phi_{b}|$, with $\Gamma_{\varphi R}\sim4.5\:\Gamma_{\varphi E}$
(see Fig. 3), consistent with dephasing caused by flux-noise. Indeed,
assuming a flux-noise power spectral density $S_{\Phi}(\omega)=A_{\Phi}/\omega$,
one can show \cite{Ithier,Yoshihara} that $\Gamma_{\varphi E}=\sqrt{A_{\Phi}\ln2}\,\left|\partial\omega_{01}/\partial\Phi_{b}\right|$
and $\Gamma_{\varphi R}=\sqrt{A_{\Phi}\ln(1/\omega_{IR}t)}\,\left|\partial\omega_{01}/\partial\Phi_{b}\right|$,
with $\omega_{IR}$ an infrared cut-off frequency determined by the
rate of data acquisition, and $\left|\partial\omega_{01}/\partial\Phi_{b}\right|\simeq2\, I_{P}\left|\varepsilon\right|/\hbar\Delta$.
In our experiments, $\sqrt{\ln(1/\omega_{IR}t)}\sim3.7$ predicting
$\Gamma_{\varphi R}\sim4.5\:\Gamma_{\varphi E}$ in agreement with
the measured value. We find an amplitude $A_{\Phi}=(2.5\:\mu\Phi_{0})^{2}$
comparable to previously reported values \cite{Yoshihara,Bylander}.

\begin{figure}
\includegraphics[width=1\columnwidth]{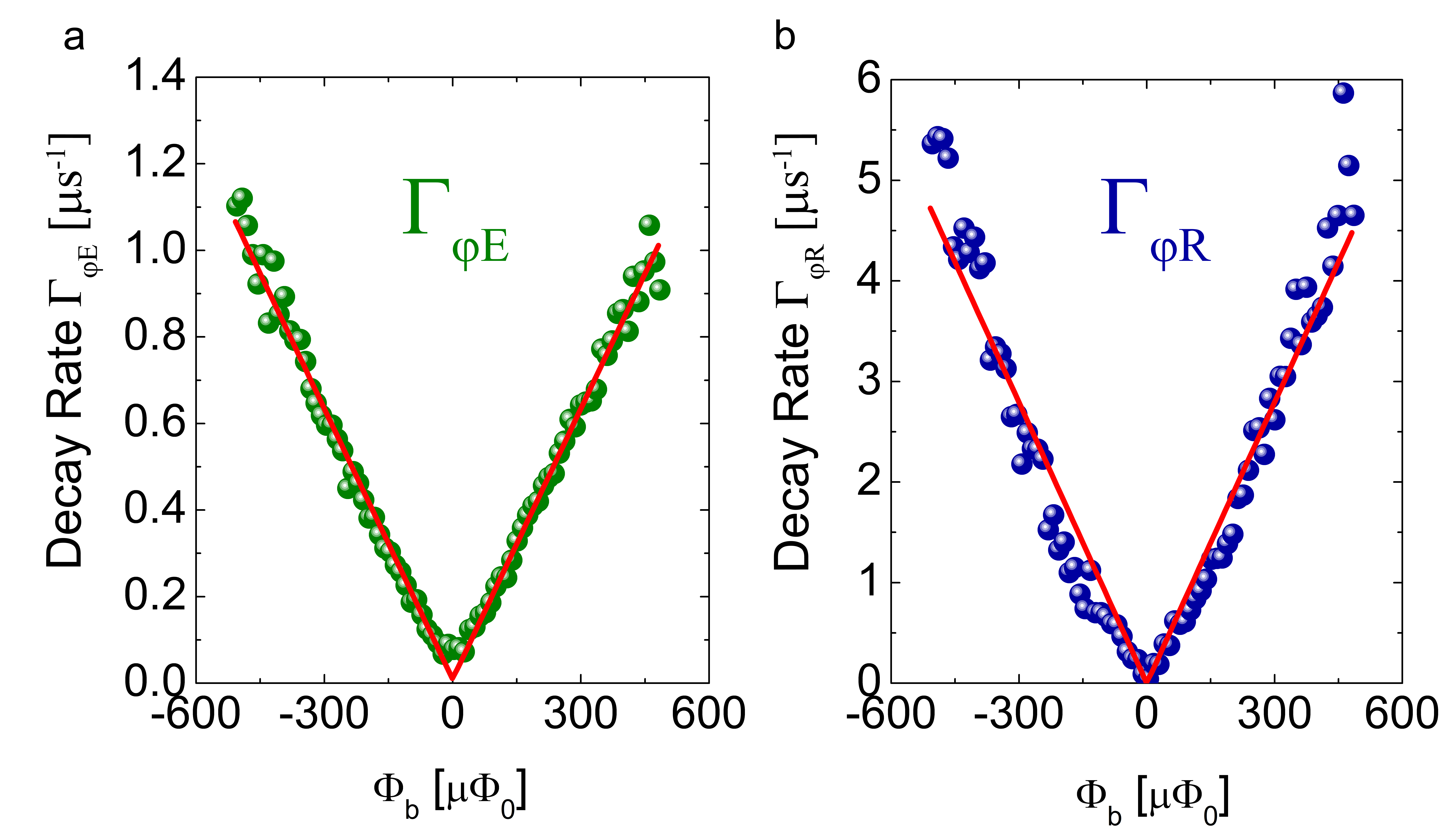}\caption{Pure dephasing rates of Q4 as a function of $\Phi_{b}$: Experimental
(dots) and fitted (line, see text) echo (a) and Ramsey (b) dephasing
rates.}
\end{figure}

All six qubits were characterized in this way, over several cooldowns
(see Table I). Energy relaxation times were found to change from cooldown
to cooldown, and also to occasionally change abruptly in the course
of one cooldown. Several mechanisms contribute to relaxation; among
them, spontaneous emission of a photon by the qubit in the measurement
lines (the so-called Purcell regime \cite{Houck}). Because this spontaneous
emission rate $T_{P}^{-1}$ is also the rate at which a photon coming
from the measurement line is absorbed by the qubit, it can be quantitatively
determined by measuring the qubit Rabi frequency $\Omega_{R}$ for
a given microwave power $P_{in}$ at the cavity input. For a qubit
coupled symmetrically to the input and output lines, one gets 
\begin{equation}
T_{P}=\frac{2}{\Omega_{R}^{2}}\frac{P_{in}}{\hbar\omega_{01}}\,.
\end{equation}
Comparing these estimates with the measured $T_{1}$ times (see Table
I), we find that $Q2$ and $Q6$ are almost Purcell-limited. The intrinsic
energy relaxation time $1/(T_{1}^{-1}-T_{P}^{-1})$ of all six measured
qubits at their optimal point is thus in the $6-20\,\mu s$ range.
This significant improvement over previous reports of $T_{1}$ in
the $0.5-4\,\mu\mathrm{s}$ range \cite{Chiorescu,Bertet,Yoshihara,Johnson}
(except one sample for which $T_{1}=12\,\mu\mathrm{s}$ \cite{Bylander})
is probably due to a combination of several factors: good control
of the electromagnetic environment in the 3D cavity \cite{Paik},
careful filtering and shielding against infrared radiation \cite{BarendsAPL},
low-loss sapphire substrate and different fabrication process. 

The frequency dependence of the relaxation rate $\Gamma_{1}=(T_{1})^{-1}$
of $Q4$ in the vicinity of the optimal point is shown in Fig. 4.
Large variations are observed, with in particular a reproducible increase
of the relaxation rate by a factor $2$ over $1$~MHz, as also recently
observed for a transmon qubit \cite{Barends}. No corresponding anomaly
in the Rabi frequency was observed at this point, which excludes spontaneous
emission into the measurement lines (see Eq. (1)). We attribute therefore
this peak to one resonant microscopic two-level system (TLS) weakly
coupled to the qubit \cite{Barends}. The remaining constant background
$\sim(20\,$$\mu\mathrm{s}){}^{-1}$ could be due to dielectric losses,
vortex motion, or out-of-equilibrium quasiparticles. 

To estimate the quasiparticle contribution to relaxation, the same
measurements were performed at $150$~mK, a temperature at which
the quasiparticle density is expected to be close to its thermal equilibrium
value. The relaxation rate increases to $\Gamma_{1}^{(150\,\mathrm{mK})}\simeq(5$~$\mu\mathrm{s})^{-1}$
due to quasiparticles with a similar frequency dependence as at $33$
mK, although less prononced. Assuming that quasiparticles are mainly
generated in the pads of the LC resonator and diffuse into the galvanically
coupled qubits, we estimate the density of quasiparticles in the vicinity
of the qubit $n_{qp}(150\,\mathrm{mK})=1\,\mu\mathrm{m}^{-3}$ \cite{Catelani}.
This density yields a theoretical relaxation rate $\Gamma_{1}^{(qp)}=(14\,\mu\mathrm{s})^{-1}$\cite{Supplementary},
lower than the measured value by a factor 3, a discrepancy which we
attribute to the crudeness of the modelling of quasiparticle diffusion.
Since $\Gamma_{1}^{(qp)}$ is proportional to the quasiparticle density
\cite{Catelani}, we conclude that an out-of-equilibrium quasi-particle
density of $\sim0.3\,\mu\mathrm{m}^{-3}$ would be sufficient to explain
the measured energy relaxation times at $33$mK, which seems a plausible
value in view of earlier reports in other superconducting qubit circuits
\cite{Paik,deVisser,Rist=0000E9}. However, the dielectric loss contribution
is also important. Taking into account reported values of dielectric
loss tangents $\sim2\times10^{-5}$ \cite{Supplementary}, we find
$\Gamma_{1}^{(dielectric)}\sim(25\,\mu\mathrm{\mathrm{\mathrm{s}})}^{-1}$,
which is comparable to the measured values. Along these lines, we
note that flux qubits fabricated on a high resistivity silicon chip
and measured with the same setup at $33$ mK showed a five-fold increase
in relaxation rate.

Another interesting aspect of our experiment is the long Ramsey pure
dephasing time $T_{\varphi R}$ measured for $Q3$ and $Q4$ at their
flux optimal point. Although not quite as long, all measured qubits
have $T_{\varphi R}$ in excess of $3\,\mu\mathrm{s}$ (see Table
I). The origin of decoherence at the optimal point for FQs has so
far not been identified. One striking feature is the large variability
of reported values of $T_{\varphi R}$ at the optimal point for rather
similar FQ samples, ranging from $0.2\,\mu\mathrm{s}$ \cite{Bertet}
up to $10\,\mu\mathrm{\mathrm{s}}$ in this work, whereas $T_{\varphi R}=2.5\,\mu\mathrm{s}$
in \cite{Bylander}. A doublet structure in the qubit line was frequently
observed at the optimal point \cite{Bertet,Yoshihara,Bal}, with a
greatly varying splitting ($20$~MHz in \cite{Bertet} and $150$~kHz
in this work, as seen in Fig. 2b) which was also found to vary in
time. 

\begin{figure}
\includegraphics[width=1\columnwidth]{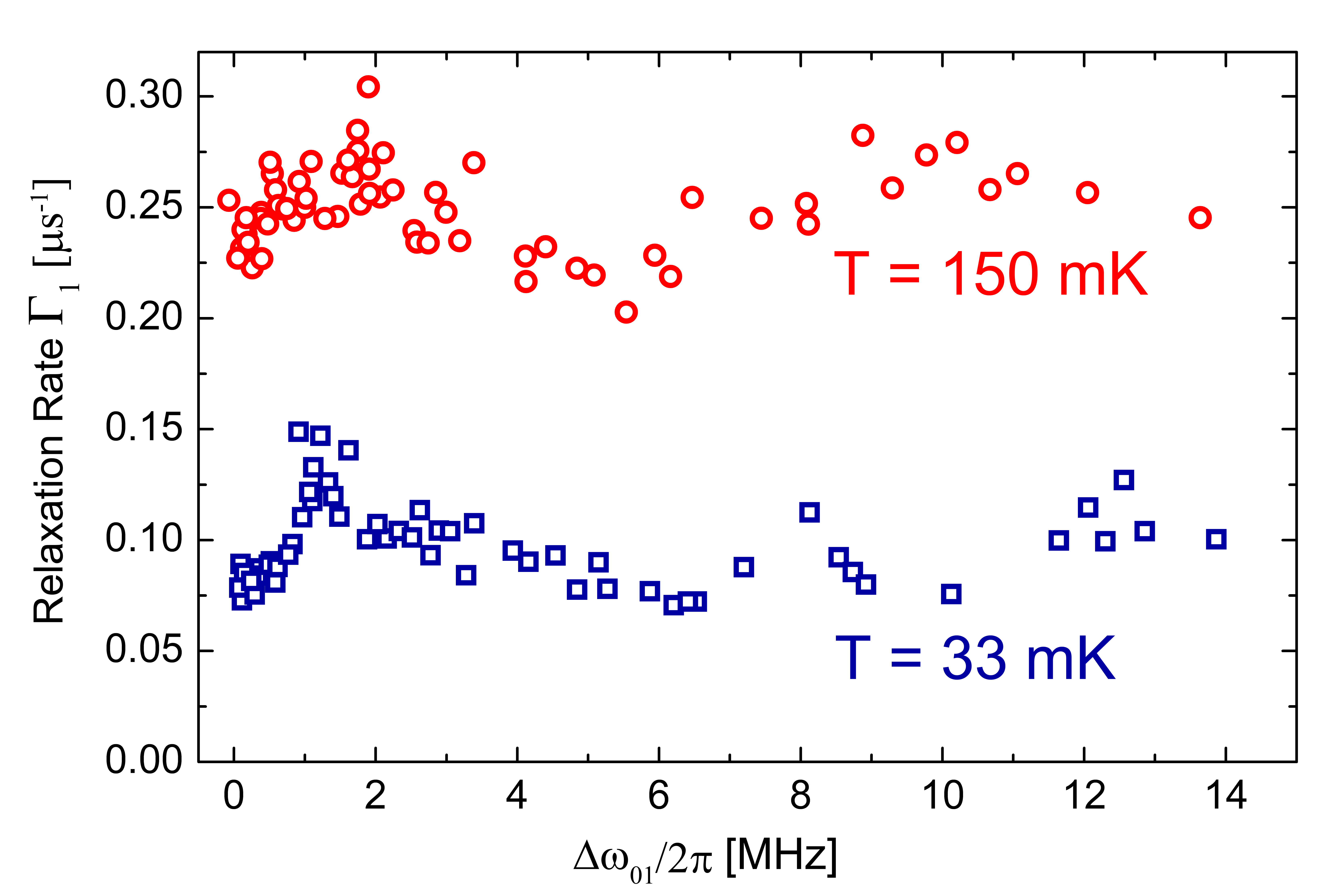}\caption{Frequency dependence of the relaxation rate $\Gamma_{1}$ of Q4 in
the vicinity of its optimal point at $T=33$mK (blue squares) and
$T=150$mK (red circles), showing an increased relaxation rate caused
by quasi-particles. }
\end{figure}

All these features are consistent with charge noise being the dominant
noise source limiting $T_{\varphi R}$ at the optimal point. The sensitivity
to this noise is exponentially dependent on the ratio $E_{J}/E_{C}$
\cite{Orlando} which is twice smaller in \cite{Bertet} compared
to the present work. Using the model described in \cite{Supplementary},
we estimate a charge modulation amplitude of $\sim50$ kHz for Q3/Q4
compared to $\sim120$ MHz for \cite{Bertet} ($\sim300$ kHz for
\cite{Bylander}), yielding a three orders of magnitude lower charge
noise sensitivity which explains qualitatively the difference in dephasing
time. The doublet lineshape of Q4 can be attributed to slow fluctuations
of the electron number parity on one of the qubit islands \cite{Bal}
as observed for transmon qubits \cite{Schreier,Rist=0000E9}.

In conclusion, we have characterized the coherence properties of six
FQs in a three-dimensional microwave cavity. We consistently find
intrinsic energy relaxation times $T_{1}$ ranging between $6$ and
$20\,\mu\mathrm{s}$, a significant improvement over previous FQ measurements
that we attribute to good control of the electromagnetic environment
provided by the 3D cavity, low-loss substrate, and careful filtering.
We identify weakly coupled two-level systems, quasiparticles and dielectric
losses as likely sources of energy damping. At the optimal point,
long Ramsey pure dephasing times up to $10\,\mu s$ are measured,
limited by a combination of photon noise and charge noise with roughly
equal contribution. We argue that charge noise is the dominant microscopic
dephasing mechanism for FQs at the optimal point, and that its effect
can be greatly minimized by chosing proper qubit parameters. Our results
prove that FQs can reliably reach long coherence times, which opens
new perspectives for the field of hybrid quantum circuits, in particular
for the coherent coupling of single spins to superconducting circuits.
\begin{acknowledgments}
We would like to acknowledge fruitful discussions with M. Devoret,
A. Lupascu and within the Quantronics group. We thank P. Sénat, P.-F.
Orfila, D. Duet, J.-C. Tack, P. Pari, D. Bouville, P. Forget and M.
de Combarieu for their technical support. This work was supported
by the ANR CHIST-ERA project QINVC, the C'Nano IdF project QUANTROCRYO,
the ERC project CIRQUSS, the JSPS and by the EU under REA Grant Agreement
No. CIG-618258 (GC).\end{acknowledgments}

\onecolumngrid
\appendix
\newpage
\part*{Supplementary Materials}

\section{Hamiltonian of the 4-junction flux qubit}

In this section, we present a model of a 4-junction flux qubit taking
into account the main geometric capacitance terms and compare the
calculation results with our experimental data. 

Fig. 5a presents the circuit diagram of our model. The loop of the
qubit is formed by four islands $I_{i=1,4}$ shown on Fig. 5b and
connected by 4 Josephson junctions. Three junctions are identical
while the area of the junction between islands $I_{1}$ and $I_{4}$
is smaller than the others by a factor $\alpha$. 

The full $(4\times4)$ geometric capacitance matrix of the circuit
was calculated using a three-dimensional electromagnetic simulator
\cite{CST}. We include in the diagram three capacitors which account
for the main contribution of the geometric capacitance. Capacitor
$C_{c}$ represents the capacitance between the largest qubit island
$I_{1}$ and $I_{3}$; the latter is connected galvanically to the
pads of the LC resonator. Capacitors $C_{g1}$ and $C_{g3}$ are respectively
the capacitances to ground of $I_{1}$ and $I_{3}$.

For qubit Q4, the values of these capacitances were found to be $C_{g1}=0.04$
fF, $C_{g3}=477$ fF and $C_{C}=0.57$ fF.

In the limit $C_{g3}\gg C_{g1}$ relevant to our experiment, the circuit
Hamiltonian can be written as: 

\begin{eqnarray}
H & = & -E_{J}\left[\cos(\varphi_{1})+\cos(\varphi_{2})+\cos(\varphi_{3})+\alpha\,\cos(\varphi_{1}+\varphi_{2}+\varphi_{3}-2\pi\Phi/\Phi_{0})\right]\label{H}\\
 &  & +4E_{C}\frac{\left(1+2\alpha+2c_{r}\right)n_{1}^{2}+\left(1+2\alpha+c_{r}\left(1+\alpha\right)\right)\left(n_{2}^{2}+n_{3}^{2}\right)-2\alpha\left(n_{1}n_{2}+n_{1}n_{3}+n_{2}n_{3}\right)-2c_{r}\left(1+\alpha\right)n_{2}n_{3}}{1+3\alpha+2c_{r}(1+\alpha)}\nonumber 
\end{eqnarray}
where $\varphi_{i}$ are the phase differences across the three large
junctions of Josephson energy $E_{J}$, the operators $n_{i}$ are
conjugated to $\varphi_{i}$ and count the number of Cooper pairs
tunneling across the junctions, $C$ is the capacitance of the large
junctions, $E_{C}$ is given by $E_{C}=\frac{e^{2}}{2C}$, $c_{r}$
is a dimensionless parameter given by $c_{r}=\frac{C_{c}+C_{g1}}{C}$,
$\Phi$ is the magnetic flux through the qubit loop, and $\Phi_{0}$
is the magnetic flux quantum. 

A complete description of the system includes all the inter-island
capacitances, the capacitances to ground of the two small qubit islands
and the inductance of the qubit loop. Assuming symmetry between the
small islands, it can be shown~\cite{Catelani-1} that the main effect
of the additional capacitances is to slightly renormalize the parameters
$E_{C}$, $\alpha$, and $c_{r}$ entering the charging energy part
of Eq.~(A1). In our numerical calculations we take this renormalization
into account, using for the geometric capacitances the values extracted
from the full capacitance matrix of our system but neglected the effect
of the loop inductance.

We approximately solve the Schrödinger equation in the charge basis
by truncating the Hilbert space to a large but finite number of charge
states ($\sim25^{3}$) and compare in Figs. c and d the results of
the calculation to the measured values of the flux qubit gap $\Delta$
and persistent current $I_{P}$. 

The Josephson energy $E_{J}$ was estimated from Ambegaokar-Baratoff
relation $E_{J}=h\Delta_{0}/(8e^{2}R_{N})$ where $\Delta_{0}$ is
the superconducting gap of a 25-nm-thin aluminium film and $R_{N}$
the tunneling resistance of the junction at low temperature and in
the normal state. Assuming $\Delta_{0}=200\:\mathrm{\mathrm{\mu eV}}$
\cite{Aumentado} and measuring the room temperature tunneling resistance
of the junctions, we find $E_{J}/h=250\,\pm15$ GHz. 

The charging energy of the junctions $E_{C}$ was estimated by measuring
their size and using standard values given in the literature for the
capacitance per unit area. Assuming $c=90\:\pm10\:\mathrm{fF/\mu m^{2}}$
\cite{Geerligs,Bouchiat,Bretheau}, we find $C=5.4\,\pm0.6$ fF and
$E_{C}/h=3.6\,\pm0.4$ GHz.

The experimental data match well with the calculation and the major
influence of adding the geometric capacitance terms is to reduce the
flux qubit gap $\Delta$ by around 1 GHz.

\begin{figure}
\includegraphics[width=1\textwidth]{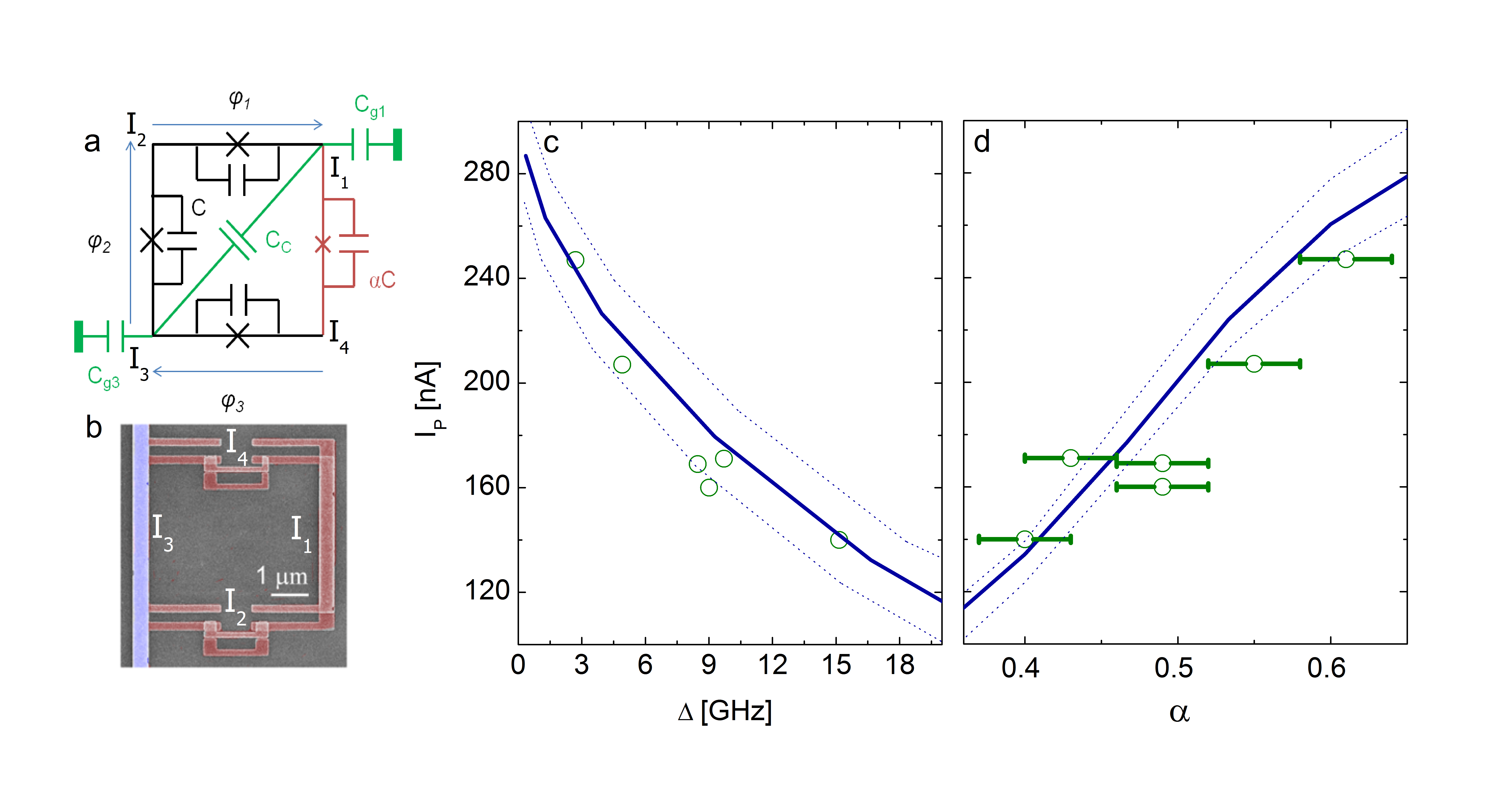}\caption{(a) Circuit diagram of the flux qubit including the main contributions
of the geometric capacitance. (b) SEM micrograph of one of the flux
qubits showing the different islands. The island $I_{3}$ (in blue)
is connected galvanically to the pads of the LC resonator. (c) Persistent
current $I_{P}$ vs its gap $\Delta$. For the blue solide line, the
parameters of the model were $E_{J}/h=250$ GHz, $E_{C}/h=3.6$ GHz,
$c_{r}=0.1102$. The upper (lower) blue dotted line corresponds to
$E_{J}/h=265$ GHz and $E_{C}/h=4$ GHz ($E_{J}/h=235$ GHz and $E_{C}/h=3.2$
GHz). The green circles are the experimental values. (d) Persistent
current $I_{P}$ vs $\alpha$ for the same parameters as (c). The
horizontal error bars correspond to the confidence interval in the
measure of $\alpha$ by SEM observation.}
\end{figure}

\section{Relaxation Rate Calculations}

In this section, we discuss the calculations of the relaxation rates
due to quasiparticles and dielectric losses; more details will be
presented elsewhere~\cite{Catelani-1}. 

Our estimate of the contribution $\Gamma_{1}^{(qp)}$ of quasiparticle
tunneling to the relaxation rate of Q4 is based on the general theory
of quasiparticle effects developed in Refs.~\cite{Catelani-2,Catelani-3}.
It was shown there that in a multi-junction qubit, $\Gamma_{1}^{(qp)}$
is obtained by summing the contributions of each individual junction,
and that at low temperatures each term in the sum is proportional
to the density of quasiparticles. A number of experiments \cite{Paik,deVisser,Rist=0000E9}
have shown that in aluminum devices at temperatures $\gtrsim150$~mK,
quasiparticles are in thermal equilibrium.

Based on the geometry of Q4, we estimate that at this temperature
there are on average just 0.1 thermal quasiparticles in island $I_{1}$.
This means that the with high probability qubit decay can only be
induced by a quasiparticle tunneling from the LC inductor gavanically
connected to the qubit into one of the small island. Therefore, indicating
with $x_{qp}=n_{qp}/2\nu_{0}\Delta_{0}$ the normalized quasiparticle
density in the inductor ($\nu_{0}$ is the aluminum density of states
at the Fermi energy and $\Delta_{0}=200\:\mu\mathrm{eV}$), we have
\begin{equation}
\Gamma_{1}^{(qp)}\simeq\frac{4}{\pi}E_{J}x_{qp}\sqrt{\frac{2\Delta_{0}}{\omega_{01}}}\left[\left|\langle1|\sin\frac{\varphi_{1}}{2}|0\rangle\right|^{2}+\left|\langle1|\sin\frac{\varphi_{2}}{2}|0\rangle\right|^{2}\right]\label{Gqp}
\end{equation}
The matrix elements can be estimated numerically after having obtained
the approximate eigenstates of the Hamiltonian in Eq.~(A1) and we
find 
\begin{equation}
\left|\langle1|\sin\frac{\varphi_{1}}{2}|0\rangle\right|\simeq\left|\langle1|\sin\frac{\varphi_{2}}{2}|0\rangle\right|\simeq0.21
\end{equation}
Since in thermal equilibrium we have $x_{qp}=\sqrt{2\pi T/\Delta_{0}}e^{-\Delta_{0}/T}$,
using the values $E_{J}=250$~GHz and $\omega_{01}=9.01$~GHz, from
Eq.~(B1) we arrive at the value $\Gamma_{1}^{(qp)}=0.07~\mu$s$^{-1}$
reported in the main text. This calculation neglects completely the
role of quasiparticles having tunneled into the other islands of the
qubit and therefore underestimates $\Gamma_{1}^{(qp)}$. 

Turning now our attention to the dielectric loss relaxation mechanism,
we use Fermi's golden rule and the quantum fluctuation-dissipation
relation to write each capacitor's $C_{i}$ contribution to the decay
rate in the form~\cite{Catelani-1} 
\begin{equation}
\Gamma_{1}^{C_{i}}=16\,\tan\delta_{i}E_{C_{i}}\sum_{j,k=1}^{3}A_{ji}A_{ki}N_{jk}\label{Gc}
\end{equation}
where $\tan\delta_{i}$ is the inverse quality factor of the material
causing the loss, the Hermitian matrix $N_{jk}$ is defined, using
the matrix elements of the number operators, as 
\begin{equation}
N_{jk}=\langle1|n_{j}|0\rangle\langle0|n_{k}|1\rangle
\end{equation}
and the dimensionless matrix $A_{ij}$ accounts for the different
coupling strengths between voltage fluctuations $V_{i}$ in each capacitors
and the number operators; specifically, this coupling adds the term
\begin{equation}
2e\sum_{i,j}n_{j}A_{ji}V_{i}
\end{equation}
to the Hamiltonian in Eq.~(A1).

For a simple order-of-magnitude estimate, we take the entries of matrix
$N_{ik}$ to be approximately identical 
\begin{equation}
N_{ik}\approx0.09
\end{equation}
with the numerical value calculated numerically using the (approximate)
eigenstates of the Hamiltonian in Eq.~(1). Then summing over indices
$j$ and $k$ and over all four junctions, we find for the junction
capacitors 
\begin{equation}
\sum_{i\in J}E_{C_{i}}\sum_{j,k=1}^{3}A_{ij}A_{ik}\simeq E_{C}\frac{(1+2c_{r})^{2}+2+\alpha(3+c_{r})^{2}}{\left[1+3\alpha+c_{r}(1+\alpha)\right]^{2}}\label{GJ}
\end{equation}
Here we neglect for simplicity the renormalization of parameters,
discussed above, due to the geometrical capacitances not included
in the diagram of Fig.~a. Similarly, summing over the geometric capacitors
we get 
\begin{equation}
\sum_{i\in G}E_{C_{i}}\sum_{j,k=1}^{3}A_{ij}A_{ik}\simeq E_{C}\frac{4\: c_{r}}{\left[1+3\alpha+c_{r}(1+\alpha)\right]^{2}}\label{GG}
\end{equation}

From the measured decay rate at 33~mK, we can put upper bounds on
the loss tangents for junction and ground capacitors; using Eqs.~(\ref{Gc})
and (\ref{GJ})-(\ref{GG}). Assuming that the relaxation of the qubit
would be due mainly to dielectric losses, we find 
\begin{equation}
\tan\delta_{J}\sim2.1\times10^{-6}\,,
\end{equation}
and 
\begin{equation}
\tan\delta_{G}\sim2.4\times10^{-5}\,.
\end{equation}
The value in Eq.(10) is an order of magnitude larger than recent estimates
\cite{Pop}, where $\tan\delta_{J}\leq3\times10^{-7}$ excluding the
junction dielectric loss as the likely source of relaxation in our
qubits. On the contrary, the number in Eq.(11) is close the the bound
$\tan\delta_{G}\leq2\times10^{-5}$ of Ref. \cite{O'Connell}, indicating
that dielectric losses in the substrate or interface oxides can be
one of the main sources of low-temperature relaxation. We note that
similar experiments, realized in the same conditions but on a high-resistivity
silicon chip, yielded an increased relaxation rate by a factor of
$\sim5$.

\end{document}